\documentclass{article}
\usepackage{a4wide}
\usepackage{ifthen}

\newboolean{anonymous}
\setboolean{anonymous}{false}
\newcommand{\ifanon}[2]{\ifthenelse{\boolean{anonymous}}{#1}{#2}}

\newboolean{smallcolumns}
\setboolean{smallcolumns}{false}
\newcommand{\ifsmall}[2]{\ifthenelse{\boolean{smallcolumns}}{#1}{#2}}

\newboolean{savespace}
\setboolean{savespace}{false}
\newcommand{\ifsaves}[2]{\ifthenelse{\boolean{savespace}}{#1}{#2}}

\newboolean{acmcolumns}
\setboolean{acmcolumns}{false}
\newcommand{\ifacmver}[2]{\ifthenelse{\boolean{acmcolumns}}{#1}{#2}}

\newboolean{pagenumbers}
\setboolean{pagenumbers}{true}
\newcommand{\haspagenumbers}[2]{\ifthenelse{\boolean{pagenumbers}}{#1}{#2}}

\usepackage{cite}
\usepackage{amsmath}
\ifacmver{}{
  \usepackage{amsthm}
}
\usepackage{xspace}
\usepackage{url}

\begin{document}


\title{On the Protocol Composition Logic PCL%
\thanks{This work was supported by the Hasler Foundation, ManCom project
2071.}
\\
{\bf Manuscript}}
\author{Cas Cremers%
\\[1ex]
{\small Department of Computer Science, ETH Zurich, 8092 Zurich, Switzerland}\\
{\small \sf{cas.cremers@inf.ethz.ch}}
}

\maketitle



\newtheorem{theorem}{Theorem}
\newtheorem{lemma}{Lemma}
\newtheorem{definition}{Definition}

\haspagenumbers{
\newcommand{\myref}[1]{on page~\pageref{#1}}
}{
\newcommand{\myref}[1]{in Section~\ref{#1}}
}

\ifacmver{
\newcommand{\mypar}[1]{\paragraph{#1} }
\newcommand{\prooftext}{}
}{
\newcommand{\mypar}[1]{\paragraph{#1.}}
\newcommand{\prooftext}{Proof}
}

\newcommand{\basicpcl}{\cite{pcljournal,thesis:datta,thesis:derek,pclentcs}\xspace}
\newcommand{\mypcl}{\cite{pcljournal,thesis:datta,thesis:derek,pclentcs,pcltls,thesis:he}\xspace}

\newcommand{\z}[1]{\mbox{#1}}
\newcommand{\zt}[1]{\mbox{{\tt #1}}}
\newcommand{\zi}[1]{\mbox{{\it #1}}}
\newcommand{\zs}[1]{\mbox{{\sf #1}}}
\newcommand{\zb}[1]{\mbox{{\bf #1}}}
\newcommand{\myhat}[1]{\ensuremath{\hat{#1}}}
\newcommand{\hx}{\myhat{X}}
\newcommand{\hy}{\myhat{Y}}
\newcommand{\lenc}{\{\mkern-4.0mu\lvert\,}				
\newcommand{\renc}{\,\rvert\mkern-4.0mu\}}				
\newcommand{\crypt}[1]{\lenc {#1} \renc}
\newcommand{\oldaxiom}[1]{\z{{\bf #1}}}
\newcommand{\newaxiom}[1]{\z{{\it(new axiom)} \; \oldaxiom{#1}}}
\newcommand{\newrule}[1]{\z{{\it(new rule)} \; \oldaxiom{#1}}}
\newcommand{\lrule}[3]{{\z{{\bf #1}}\;\;}\frac{#2}{#3}} 
\newcommand{\emptyseq}{[]}

\newcommand{\HONEST}{\zi{HONEST}}
\newcommand{\runsof}{\zs{Runs}}

\newcommand{\za}[1]{\zt{#1\;}}
\newcommand{\aaction}{\za{a}}
\newcommand{\asend}{\za{send}}
\newcommand{\areceive}{\za{receive}}
\newcommand{\anew}{\za{new}}
\newcommand{\amatch}{\za{match}}
\newcommand{\aencrypt}{\za{enc}}
\newcommand{\adecrypt}{\za{dec}}
\newcommand{\averify}{\za{verify}}
\newcommand{\asign}{\za{sign}}

\newcommand{\paction}{\zs{a}}
\newcommand{\Honest}{\zs{Honest}}
\newcommand{\psend}{\zs{Send}}
\newcommand{\preceive}{\zs{Receive}}
\newcommand{\pgen}{\zs{Gen}}
\newcommand{\pencrypt}{\zs{Encrypt}}
\newcommand{\pdecrypt}{\zs{Decrypt}}
\newcommand{\pverify}{\zs{Verify}}
\newcommand{\pcontains}{\zs{Contains}}
\newcommand{\phas}{\zs{Has}}
\newcommand{\pcomputes}{\zs{Computes}}
\newcommand{\pfresh}{\zs{Fresh}}

\newcommand{\thash}{\zi{HASH}}
\newcommand{\tenc}{\zi{ENC}}
\newcommand{\tsig}{\zi{SIG}}

\newcommand{\PKINIT}{\mathrm{PKINIT}}
\newcommand{\CR}{\mathrm{CR}}
\newcommand{\crinit}{\ensuremath{\z{Init}_{\CR}}}
\newcommand{\crresp}{\ensuremath{\z{Resp}_{\CR}}}
\newcommand{\BS}{\mathrm{BS}}

\newcommand{\remark}[2]{

\noindent
\framebox[\textwidth][c]{
\parbox{0.97\textwidth}{\noindent\small\textsf{{\bf #1:} #2}}}
\smallskip

}

\newcommand{\weakauth}{\phi_{\mathrm{weak-auth}}}
\newcommand{\strongauth}{\phi_{\mathrm{auth}}}
\newcommand{\formauth}{\Phi_{\mathrm{weak-auth}}}
\newcommand{\formexpandauth}{\top [ \; \crinit \; ]_X \Honest(\hy) \land \hy \not= \hx \supset \weakauth}
\newcommand{\myrun}{R_{\mathrm{no-auth}}}


\begin{abstract}
\noindent
A recent development in formal security protocol analysis is the
Protocol Composition Logic (PCL).  We identify a number of problems with
this logic as well as with extensions of the logic, as defined
in~\cite{pcljournal,pcltls,thesis:he,thesis:datta,thesis:derek,pclentcs}.
The identified problems imply strong restrictions on the scope of PCL,
and imply that some currently claimed PCL proofs cannot be proven within
the logic, or make use of unsound axioms. Where possible, we propose
solutions for these problems.

\end{abstract}



\section{Introduction}

Formally establishing properties of security protocols has been
investigated extensively during the last twenty years, but a definite
model and a corresponding method for security protocol analysis has
remained elusive thus far.

The most successful approaches to security protocol analysis have been
focused on tools for finding attacks, which are often based on bounded
model checking or constraint solving,
e.g.~\cite{lowe97casper,avispatool}. When such tools find an attack, one
can easily verify manually whether or not the attack actually
exists on the protocol. Some tools even allow for unbounded verification
of protocols, e.g.~\cite{proverif,scytherthesis}. If no attack is found
with such tools, correctness of the protocol follows.
However, this provides little insight into why a protocol is
correct.

An alternative approach is to develop a logic for reasoning about
security protocols. When a protocol is proven correct in such a logic, the
derivation steps can provide insight into the mechanisms that make the
protocol work. Despite the obvious promise of such an approach, several
attempts have failed, most notably the BAN logic from \cite{ban90}. One
of the stumbling points seems to be to provide a logic that is sound
with respect to the complicated semantics of security protocol execution
in the presence of an active intruder, and is able
to provide concise formal proofs.

A recent attempt to develop such a logic is the Protocol
Composition Logic (PCL) from e.g.~\cite{pclentcs}. This logic has evolved
from a protocol model to express protocol composition and
refinement, into a model with an associated logic that can be used to
formally prove security properties of
protocols \cite{pcljournal,thesis:datta,thesis:derek}. 
Variants of PCL have been applied to many case studies and offer
several interesting features. For example, one can reason about security
protocols without explicitly reasoning about the (complex) intruder by means
of a special kind of invariant reasoning captured by the {\em honesty rule}.
This kind of reasoning also allows the protocol logic to deal with protocol composition
and refinement, where proofs can be reused. PCL has been extended with
several features in further work, such as an extension for hash
functions in \cite{pcltls} that was used for a modular correctness proof of IEEE
802.11i and TLS.

In this paper, we identify a number of problems with PCL as defined
in \mypcl. They
have implications for the scope of PCL, a number of claimed formal
proofs, and several extensions to the base model.  In particular, we show that in contrast with the claims in e.g.~the introduction of \cite{pclentcs}, 
PCL as defined in \basicpcl cannot be used to prove common
authentication properties of protocols that do not include signatures.
We show that a number of claimed proofs in PCL cannot be
correct because (a) there is no way to establish preceding actions in a
thread, and (b) there is no way to express type restrictions in PCL.
With respect to existing PCL extensions, we identify two problems: the
Diffie-Hellman extension from \basicpcl does not correctly capture the algebraic behaviour of
Diffie-Hellman-like protocols, and the extension for hash functions from \cite{pcltls,thesis:he} is
not sound.  Some of these problems can be resolved by minor modifications
to PCL, but other problems require further investigation. Our observations
suggest that it is at least required to make changes to existing axioms,
to introduce new axioms, and to add a mechanism for a type system.

The purpose of this paper is to identify some of the challenges that
need to be addressed in order to make a logic like PCL work. We hope it
will contribute to the improvement of PCL, and will lead to a better
understanding of some of the pitfalls of designing a compact and usable
formal logic for security protocols.

{\bf The scope of this paper.}
The presentation of this paper is inherently difficult, not least
because there are a number of different papers on PCL, which
vary in notation and technical details.  Many ideas
were already present in precursors of PCL,
e.g.~\cite{pclroots,pcllogic}, but these variants use different concepts
than later versions of PCL. These early variants
in~\cite{pclroots,pcllogic} have no notion of thread (a.k.a.~process,
run, or role instance), and events are bound to agents.  More recent
versions of PCL bind events to threads of agents, and therefore
distinguish between several threads of the same agent. PCL versions of
the latter type can be found in \cite{ddmp03,ddmp03b,ddmp04b,csfw04}.
Subsequently, \cite{ddmp03,ddmp03b,ddmp04b,csfw04} have been claimed to
be either subsumed, or revised and extended, by more recent works
\basicpcl.  Hence we choose here to focus on \basicpcl, which contain
similar descriptions of PCL.  Throughout this paper we write {\em basic
PCL} to refer to \basicpcl.  The publications on basic PCL describe the
fundamental part of PCL that focusses on authentication.  In general,
the comments in this paper apply to basic PCL.  The comments in
Section~\ref{sec:hashaxioms} apply only to the extensions found in
\cite{pcltls,thesis:he}. Our comments here do not cover the recent
extensions to basic PCL for the analysis of secrecy, as found in
\cite{pclsecrecy}, nor the computational variants of PCL, as found in
e.g.~\cite{csfw06}. 

{\bf Syntax and page references.}
In order to pinpoint our observations to specific formulas, we select a
specific version of PCL to refer to.  We have chosen the most recent
description of PCL from 2007 as found in \cite{pclentcs}.  
In particular, we will use \cite{pclentcs} as a reference for the syntax of PCL
formulas, and to provide specific page references.
Hence we use \cite{pclentcs} as the reference paper to present the
problems with basic PCL from the papers \basicpcl. 

For the technical details, in particular the formulas, we assume the
reader is at least somewhat familiar with one of the papers
from \basicpcl or \cite{thesis:he,pcltls}.
However, the main points should be clear to readers familiar with
formal security protocol analysis.

The remainder of the paper is structured in the following way. We start
off by recalling some PCL notation and concepts in
Section~\ref{sec:prelim}. Then, in Section~\ref{sec:basicpcl} we discuss
problems with the basic definition of PCL. In Section~\ref{sec:pclextensions} we
identify two problems with existing PCL extensions. We conclude in
Section~\ref{sec:conclusions}.

\mypar{Acknowledgements}
The author would like to thank
David Basin, 
Anupam Datta,
Felix Klaedtke,
Sjouke Mauw,
Simon Meier,
John C. Mitchell,
Arnab Roy
and the anonymous referees
for useful discussions and feedback on earlier
versions of this paper.

\section{Preliminaries}
\label{sec:prelim}

The purpose of this section is to recall some PCL notions 
 that are required to interpret the
forthcoming sections. We use the syntax from \cite{pclentcs}. This partial summary of PCL is incomplete, and
we encourage the reader to use the original papers for reference.

The structure of PCL is as follows: first notation is introduced to
define terms, which in turn are used to define protocols. For such protocols, an execution model is defined,
assigning to each protocol a set of possible execution
histories, called {\em runs}.
Then, a protocol logic is defined in order to reason about (sets of) runs
of a protocol. This logic is proven sound with respect to the execution
model. This means that if one proves a property in terms of the protocol
logic, such as $\preceive(\ldots,m)$, then a similar property  should hold for the corresponding set of runs
in the execution model, such as ``$\areceive \ldots,m$ has occurred in the protocol run''. We touch upon these elements below.

{\bf Terms.} A term system is introduced that contains constants
(nonces, names, keys, etc.), variables, and compound terms such as
tuples, encryptions, and signatures.

{\bf Protocols.}
In PCL, a protocol $Q$ is defined as a set of roles. Each role $\rho
\in Q$ is defined as a list of actions. Examples of possible
actions can be found in the {\em actions} column of
Table~\ref{tb:examples}, and correspond respectively to sending
terms, receiving terms, generating fresh terms, encryption, decryption,
and signature verification.  Each role is partitioned into a set of {\em
basic sequences}.  A basic sequence $\BS_i$ of a role $\rho$ is a
contiguous subsequence of $\rho$ that starts with either the first
action of $\rho$ or a \areceive action, and of which the other actions are
not \areceive actions.
The basic sequences intuitively represent the idea that agents only
block at \areceive actions, and execute all other actions immediately,
allowing one to regard basic sequences as atomic actions in some
respects.
The notion of basic sequences therefore roughly corresponds to the notion
of {\em step-compression} in other models, e.g.~\cite{DBLP:conf/esorics/BasinMV03}.
The basic sequences of a protocol are defined as
the union of the basic sequences of each of its roles, and hence each
action of a role of a protocol occurs in exactly one of its basic sequences.

Observe that although some versions of PCL have two
distinct actions \areceive (for reading a term from the network without
parsing) and \amatch (for parsing a term and blocking if it is
not as expected), these two actions are in many proofs collapsed together
into a single \areceive action that receives and matches terms at
the same time.

{\bf Execution model.} An execution model is defined for protocols 
(in terms of {\em cords}). Actions are executed by {\em agents},
with names like $\hx,\hy$. A subset of these agents, defined as
$\HONEST(C)$ (where $C$ is the initial configuration of the
system), are called the {\em honest agents}, and execute their protocol
roles as expected. The agents may execute each protocol role any number
of times; each such instance of a role is called a {\em thread} (in
other formalisms, this notion is known as a strand, or a process, or a
run).
Threads are usually denoted by symbols such as $X,Y,Z$: a sequence of
actions $P$ executed within a thread $X$ is written as $[P]_X$. The
notation $\hx$ is often%
\footnote{The hat notation $\myhat{ }$ is
used in at least two different interpretations in both \cite{pcljournal}
and \cite{pclentcs}.
In some cases $X$ is used to denote a particular thread, and $\hx$ is
then interpreted as the agent that executes that thread,
as e.g.~can be seen
in the usage of $Y$ in the {\bf SEC} axiom \cite[page 327]{pclentcs},
replicated here \myref{pg:sec}. Thus, in one interpretation $\myhat{ }$
can be regarded as a function from a thread to an agent.
However, in other cases $\hx$ is used to denote a particular agent, and
$X$ is then interpreted as ``any thread executed by the agent $\hx$'', as
in 
the honesty rule {\bf HON} \cite[page 329]{pclentcs}, and
the {\bf VER} axiom \cite[pp. 329--330]{pclentcs}, replicated here 
\myref{pg:ver}.
In this second interpretation, $X$ is a variable that ranges over all threads
executed by the agent $\hx$, and $\hx$ effectively expresses a domain
restriction on $X$.
}
used to denote the agent executing the thread $X$.  Informally, if the
agent $\hx$ is honest, $[P]_X$ is a sequence of actions from a protocol
role.  The agents that are not part of $\HONEST(C)$ can execute
so-called intruder roles, in line with the common Dolev-Yao intruder
model.  In the context of this execution model, the protocol description $Q$ gives rise to a
set of {\em runs} denoted by $\runsof(Q)$.
A run is typically denoted by $R$, and corresponds to
a sequence of actions as executed by threads. A run represents a
possible execution history of the system. 

\ifacmver{%
\newcommand{\pclterms}{$m,x,t,t'$\xspace}%
}{%
\newcommand{\pclterms}{$m,x,enct,sigt,bigt,smallt$\xspace}%
}%
\begin{table}[t]
\centering
\begin{tabular}{|l|l|l|}
\hline 
Action &
Associated term &
Predicate \\
& structure & (in thread $X$) \\
\hline 
\ifacmver{%
$\asend \hx,\hy,m$		& & $\psend(X,m)$	 \\
$\areceive \hx,\hy,m$	& 	& $\preceive(X,m)$  \\
$\anew x$			& & $\pgen(X,x)$	 \\
$\aencrypt m,K$			& $t = \tenc_K\crypt{m}$ & $\pencrypt(X,t)$	\\
$\adecrypt t,K$		& 	$t = \tenc_K\crypt{m}$ & $\pdecrypt(X,t)$	\\
$\averify t,m,K$		& $t = \tsig_K\crypt{m}$ & $\pverify(X,t)$	\\
& & $\phas(X,m) $ \\
& & $\Honest(\hx) $ \\
& & $\pcontains(t,t') $ \\
}{%
$\asend \hx,\hy,m$		& & $\psend(X,m)$	 \\
$\areceive \hx,\hy,m$	& 	& $\preceive(X,m)$  \\
$\anew x$			& & $\pgen(X,x)$	 \\
$\aencrypt m,K$			& $enct = \tenc_K\crypt{m}$ & $\pencrypt(X,enct)$	\\
$\adecrypt enct,K$		& 	$enct = \tenc_K\crypt{m}$ & $\pdecrypt(X,enct)$	\\
$\averify sigt,m,K$		& $sigt = \tsig_K\crypt{m}$ & $\pverify(X,sigt)$	\\
& & $\phas(X,m) $ \\
& & $\Honest(\hx) $ \\
& & $\pcontains(bigt,smallt) $ \\
}
\hline 
\end{tabular}
\caption{Some examples of PCL actions, action predicates, terms, and their
relations. Here $\hx,\hy$ denote agents, \pclterms
denote terms, and $X$ denotes a thread.}
\label{tb:examples}
\end{table}%

{\bf Protocol logic.} For each of the actions that can occur in a run, a
corresponding {\em action predicate} is defined. Some examples are given
in the right column of Table~\ref{tb:examples}. The protocol logic is
extended with logical connectives, a predicate to reason about the
ordering of actions in a run ($<$), as well as predicates to reason about
the knowledge of threads.

One of the predicates of the logic is the $\Honest$ predicate, which is
closely related to the set $\HONEST(C)$. For a run $R$,
$\Honest(\hx)$ is used to denote that $\hx \in \HONEST(C)$ and
``all threads of $\hx$ are in a `pausing' state in $R$'' \cite[page
325]{pclentcs}. This is a technicality\footnote{The $\Honest$ predicate
serves within the honesty rule 
as an encoding of the atomic nature of 
basic sequences.} that is needed to make the
honesty rule (which is not described here) work.

One predicate of interest for this paper is the $\pcontains$ predicate,
which is used to reason about the relation between two terms.  In
particular,
$\pcontains(t_1,t_2)$ is defined by means of the subterm relation
$\subseteq$ in \cite[page 323]{pclentcs}, stating that $t_1$ contains
$t_2$ as a subterm. The subterm relation $\subseteq$ is
never formally defined.
Here we assume
that the subterm relation is defined syntactically%
\footnote{Observe that if we would alternatively assume that the subterm
relation involves only tuple projection, as seems to be suggested by the
{\bf CON} axiom \cite[page 444]{pcljournal}, i.e.~$t \subseteq t' \equiv
(t=t') \lor (\exists t''.t'=(t,t'') \lor t'=(t'',t))$, then e.g.~the
{\bf P2} and {\bf VER} axioms in \cite{pcljournal,pclentcs} are unsound,
because respectively the fresh value and the signature might be sent as
part of an encrypted term, and decrypted by an agent that knows the
key.}%
.
In particular, we assume that a
signature such as $\tsig_{\hx}\crypt{m}$ contains $m$,
i.e.~$\pcontains(\tsig_{\hx}\crypt{m},m)$ holds.
Note that this assumption only plays a role in the construction of a particular counterexample in
Section~\ref{sec:notypes}, and does not influence any of our general
observations.

Using the protocol logic, one aims to establish properties of all runs of a
protocol. A run $R$ of a protocol $Q$ that satisfies a property
$\phi$ is denoted as $Q,R \models \phi$. If all runs of a protocol $Q$
satisfy $\phi$, we write $Q \models \phi$. If a formula $\phi$ is provable
using the logic by using any PCL axioms except the honesty rule, we write
${} \vdash \phi$, which expresses that $\phi$ holds for all protocols. For
formulas provable using the axioms and the honesty rule for a protocol
$Q$, we write $\vdash_Q \phi$, which expresses that $\phi$ holds for the
protocol $Q$.

\medskip
\noindent
In the remainder of this paper we use the following convention: All
formulas that are numbered are ours. All unnumbered formulas, including
the named axioms, are
copied from PCL papers, in which case the source paper and page
number is given.

\section{Problems with basic PCL}
\label{sec:basicpcl}

In this section, we address two problems with basic PCL as defined
in \basicpcl. First, we identify
in Section~\ref{sec:authexistence} a strong restriction on the scope
of basic PCL. Then, we identify two
missing proof mechanisms that seem to be necessary to prove simple
protocols correct in Sections \ref{sec:noprec} and~\ref{sec:notypes}.

\ifacmver{
\subsection{Authentication properties only provable \\ for signature protocols}
}{
\subsection{Authentication properties only provable for signature protocols}
}

\label{sec:authexistence}

In basic PCL \basicpcl, the possibility of
proving authentication properties is limited to protocols that use
signatures. 

Proving an authentication property such as aliveness or agreement requires a
proof of existence of actions of another agent. In basic PCL, there is only one axiom
that allows for a conclusion of that type. The precondition of this axiom can
only be met by protocols that use signatures.  As a consequence, if a protocol
does not use signatures, the existence of a communication partner cannot be
proven within the logic.

This is surprising, as PCL was ``initially designed as a logic of
authentication'' \cite[page 313]{pclentcs}, and it is stated in the
abstract of the paper that ``PCL [\ldots] has been applied to a number
of industry standards including SSL/TLS, IEEE 802.11i [\ldots]''
\cite[page 311, abstract]{pclentcs}. These
protocols consist of many subprotocols that do not rely only on
signatures, but also rely on symmetric key cryptography and symmetric
MACs.

The problem occurs for all authentication properties that imply the
existence of a communication partner. This includes {\em aliveness} or any form of 
{\em agreement} from \cite{lowe97hierarchy}, 
\ifanon{or}{}
{\em matching conversations} from \cite{matchingconversations}%
\ifanon{.}{%
, or {\em synchronisation} from \cite{CMV2006synchronisation}.}
The matching conversations property is the authentication property used within basic PCL \cite[page 331]{pclentcs}.
All these properties are of the form:
\begin{equation}
\label{eq:authproperty}
  \phi(X) \; \supset \; \exists Y . \psi(Y)
\end{equation}
where typically, $\phi(X)$ denotes that thread $X$ executes a role
that authenticates another role.
Such a property states that if a thread $X$ executes
actions of a certain role (e.g.~all actions of an initiator role), then
there exists a thread $Y$ that has executed some actions of another
role (e.g.~the
first two actions of a responder role), and possibly some further
condition holds. This is captured in $\psi(Y)$. Note that both the weak and
strong authentication properties in the example of \cite[page
331]{pclentcs} belong to this class of authentication properties.

In order to prove such a property, it is required to prove the existence of a
thread. 
Examination of all axioms of the logic reveals that only one axiom
allows for the conclusion of the existence of a thread, which is the
{\bf VER} axiom \cite[page 327]{pclentcs}:
\begin{multline*}
\mbox{{\bf VER}} 
\ifacmver{\;\;}{\qquad}
\Honest(\hx) \land \pverify(Y,\tsig_{\hx}\crypt{x}) \land
\hx \not= \hy \supset \\
\exists X.\psend(X,m) \land \pcontains(m,\tsig_{\hy}\crypt{x})
\end{multline*}
\label{pg:ver}%
Because this axiom has the signature verification predicate $\pverify$ in the precondition, it can
only be used for signature protocols. Hence there is no way to prove the
existence of another thread for protocols without signatures.

\mypar{Other comments regarding non-signature protocols}

Whilst introducing new axioms for establishing thread existence for
non-signature protocols is non-trivial, there is a related problem with the
inconsistent use of symmetric/asymmetric cryptography.  In basic PCL, there is
only one type of encryption operator, and only a single key
set, in the sense that the rules for encryption and decryption do not
distinguish between different types of keys.
This suggests that either only symmetric, or only asymmetric encryption is
supported.

The definitions of the reaction rules \cite[page 318]{pclentcs},
the action formulas \cite[page 324]{pclentcs} and the
$\phas$ predicate \cite[pp.324--325]{pclentcs} all indicate that one encrypts and
decrypts a message using the same key, e.g.:
\ifacmver{%
\begin{alignat*}{2}
  \phas_{i+1}(A,\tenc_K\crypt{m}) 
  \quad
  \z{if } \;
  \phas_i(A,m)
  &
  \z{ and }
  \phas_i(A,K)
  \\
  \phas_{i+1}(A,m) 
  \quad
  \z{if } \;
  \phas_i(A,\tenc_K\crypt{m}) 
  &
  \z{ and }
  \phas_i(A,K)
\end{alignat*}
}{%
\begin{multline*}
\shoveright{
\begin{array}{ll}
  \phas_{i+1}(A,\tenc_K\crypt{m}) 
  & \z{if } \;
  \phas_i(A,m)
  \z{ and }
  \phas_i(A,K)
  \\
  \phas_{i+1}(A,m) 
  & \z{if } \;
  \phas_i(A,\tenc_K\crypt{m}) 
  \z{ and }
  \phas_i(A,K)
\end{array}
}
\end{multline*}
}
The same assumption, that one encrypts and decrypts with the same key,
can be found in axiom {\bf AR3} \cite[page 327]{pclentcs}. Without
explaining the full details of the notation of this axiom, we want to point out that the key
$K$ is used as the key to decrypt a message encrypted with $K$:
\begin{multline*}
\shoveright{
\begin{array}{ll}
\oldaxiom{AR3} 
&
\paction(x) \; [ y:= \adecrypt x,K ]_X \; \paction(\tenc_K\crypt{y})
\end{array}
}
\end{multline*}
However, the idea that
symmetric encryption is intended seems to be contradicted by the
{\bf SEC} axiom \cite[page 327]{pclentcs}, which states that there is
only one agent that can decrypt a message encrypted with a key, along
the lines of asymmetric encryption:
\ifacmver{%
\begin{alignat*}{1}
\oldaxiom{SEC}
\;\;
\Honest(\hx)
\land
\pdecrypt(Y,\tenc_{\hx}\crypt{x}) \supset (\hy = \hx)
\end{alignat*}
}{%
\begin{multline*}
\shoveright{
\ifacmver{}{\;\;\;}
\oldaxiom{SEC}
\ifacmver{\;\;}{\qquad}
\Honest(\hx)
\land
\pdecrypt(Y,\tenc_{\hx}\crypt{x}) \supset (\hy = \hx)
}
\end{multline*}
}%
\label{pg:sec}%
Combined with the definition of the $\phas$ predicate,
one is lead to conclude that the only one who can create the encrypted
message that occurs in the {\bf SEC} axiom, is the agent that can
decrypt it.

\subsubsection{Implications}

In order to prove any authentication property of the form
of~(\ref{eq:authproperty}) for a protocol that does not use signatures,
one needs to introduce consistent machinery for symmetric and asymmetric
cryptography and several new axioms.

In basic PCL, protocols like Needham-Schroeder-Lowe\footnote{Observe that in
a precursor of PCL in \cite{pcllogic} there is a proof of Needham-Schroeder-Lowe,
but this logic has only asymmetric cryptography, and has no notion of 
threads or processes.} and many key establishment protocols cannot be modeled,
and even if they could be, no authentication proofs could be given for them.
Similarly, it is impossible to use basic PCL as defined in \basicpcl to prove
the authentication properties of the protocols ``SSL/TLS, IEEE 802.11i''
\cite[page 311, abstract]{pclentcs}. In order to prove authentication of
these protocols, one is required to introduce additional axioms.

\subsubsection{Resolving the problem}

The problem can be split into three subparts. First, symmetric and asymmetric
cryptography must be dealt with consistently.  Second, basic PCL must be
extended with axioms that enable proving authentication properties
(existence of a thread) for symmetric cryptography. Third, resolving the
problem for public-key encryption protocols (which includes many key
agreement protocols, and the well-known Needham-Schroeder-Lowe protocol) requires
the introduction of additional theory. We first
address the two easier problems before turning to public-key encryption.

\mypar{Dealing consistently with symmetric and asymmetric cryptography}

The first requirement for resolving this problem is to split
symmetric and asymmetric encryption either by having two types of
encrypt/decrypt actions as in e.g.~\cite{avispatool}, or by splitting the
key set as in e.g.~\cite{scytherthesis}. Either choice impacts the action sets,
the action formulas, the $\phas$ predicate, and requires the addition of
further {\bf ARx} axioms and an alternative for the {\bf SEC} axiom. Most of
the additions are trivial, but introduce additional complexity into the
logic and possibly also the execution model.

\mypar{Proving authentication for protocols using symmetric
encryption or keyed hashes}

The axiom for signature protocols deals with the simplest possible case
of authentication by cryptography: the signature of an honest agent can be used
immediately to conclude that there exists a thread executed by this agent. If
that agent differs from any agent executing the currently known threads, this
implies the existence of another thread. This conclusion is based on the fact
that only one agent has the private key needed to perform the signature.

For symmetric encryption and keyed hashes there is usually a notion of
symmetry: in most cases, two agents share a symmetric key. Thus, if a symmetric
key $K$ shared by two honest agents $\hx$ and $\hy$,
occurs in a message of a thread $X$, there are two candidates for the agent who
created the message. If we can exclude that the message was generated in thread
$X$ executed by $\hx$, we can derive the existence of another thread $Y$ that
must have created it.

The authors of PCL have explored a variant of this route for e.g.~the
extensions needed for the keyed-hash based protocols in \cite{pcltls}.
We discuss the merits of those extensions in detail in
Section~\ref{sec:hashaxioms}.

\mypar{Proving authentication for public-key encryption protocols}

If we assume that basic PCL is extended to consistently deal with public-key
encryption, the authentication properties still cannot be proven, as we
lack an axiom for establishing thread existence as pointed out
previously.  Contrary to signatures or symmetric-key cryptography, there
is no easy solution with an axiom based on honest agents only, because
the public keys are also known to the intruder. If a message encrypted
with the public key of an honest agent $\hx$ occurs, no conclusion
can be drawn about the sender or creator of the message.  
Thus we must also consider the possibility that the message was
sent by the adversary. 

A first step towards a solution would be to
introduce an axiom in the protocol logic, along the lines of Lemma 2.3
\cite[page 321]{pclentcs} of the execution model. In
fact, such an axiom was present in a precursor of basic PCL. In \cite[page
701]{pcllogic} one can find axiom {\bf AR1} which, when recast in the
basic PCL
model, would read
\ifacmver{
\begin{multline}
  \z{{\bf AR1}} 
  \ifsmall{\; \; \;}{\quad}
  [\areceive t]_X
  \exists Z.\psend(Z,t) \\
\z{(This axiom is not in basic PCL)}
\end{multline}
}{%
\begin{multline}
\shoveright{
  \z{{\bf AR1}} 
  \ifsmall{\; \; \;}{\quad}
  [\areceive t]_X
  \exists Z.\psend(Z,t) 
  \ifsmall{\; \; \;
  \z{{\small (This axiom is not in basic PCL)}}
  }{\qquad
  \z{(This axiom is not in basic PCL)}
  }
}
\end{multline}
}
A second axiom that might be adapted for proving these properties is the {\bf SRC}
axiom in \cite[page 703]{pcllogic}.

Unfortunately, for either of these axioms, authentication proofs would
require either further machinery to prove that $Z$ is executed by an
honest agent, or explicit reasoning about the intruder, as the sender
thread $Z$ need not be executed by an honest agent. This type of
reasoning is not supported by basic PCL, and would require a significant
amount of additional machinery. 

\subsection{No means to establish preceding actions in a thread}
\label{sec:noprec}

By the definition of the PCL execution model, threads of honest agents start at
the beginning of a role description, and always execute the actions in the
order given by the role. Thus, if one can establish that a thread of an honest
agent has executed the $n$th action of a role, it should also be possible to
conclude a similar result about preceding actions within the logic: in
particular, one should be able to conclude that the preceding $n-1$ actions
have also occurred in the same thread. However, there is no mechanism
in the logic of basic PCL \basicpcl to draw such a conclusion.

Note that one can
use the honesty rule to prove from, e.g., a \asend action that a \areceive
action must have occurred previously, but only if both actions occur within the
same basic sequence. However, if one wants to prove that given a basic
sequence, an action has occurred from another basic sequence,
there are no rules to enable this type of reasoning.

\subsubsection{Implications}

The consequence of not having a means of establishing preceding actions is
that some claimed proofs do not seem to be correct. 
For example, observe the initiator role of the $Q_{\CR}$ protocol \cite[page 315]{pclentcs}. In
order to prove invariants for the protocol using the honesty rule, one has to prove that
the invariants hold for all basic sequences of the protocol. The
initiator role consists of two basic sequences $\BS_1$ and $\BS_2$
\cite[page 335]{pclentcs}:
\begin{align*}
\BS_1 \equiv [ \; & \anew m ;\;
\asend \hx,\hy,m ;\; ]_X
\\
\BS_2 \equiv [ \; & \areceive \hy,\hx,y,s ;\; \averify
s,(y,m,\hx),\hy ;\;
\ifacmver{\\ & }{
\ifsmall{\; \; \qquad \qquad \qquad \qquad { } \\&\quad \;\;\, }{}
}
r := \asign(y,m,\hy), \hx ;\; \asend
\hx,\hy,r ;\; ]_X
\end{align*}\label{pg:bs1bs2}%
In this protocol, the basic sequence $\BS_1$ defines $m$ as a generated
value, which determines the semantics of $m$ in $\BS_2$. Observe that there is
no way to tell if $m$ was received or generated from just investigating $\BS_2$. 

In order to show how this makes certain proofs impossible in basic PCL,
consider a protocol $Q'$, that contains a role $\rho$ consisting of 
the basic sequences $\BS_1'$ followed by $\BS_2$, where $\BS_1'$ is defined
as:
\ifacmver{
\begin{equation}
  \BS_1' \equiv [ \; \areceive \tenc_K\crypt{m} ;\; \asend \hx,\hy,m ;\; ]_X
\end{equation}
}{%
\begin{multline}
\shoveright{
  \BS_1' \equiv [ \areceive \tenc_K\crypt{m}; \asend \hx,\hy,m; ]_X
}
\end{multline}
}
Thus the protocol $Q'$ shares the basic sequence $\BS_2$ with $Q_{\CR}$,
but in $Q'$, $m$ is a received valued as opposed to a generated value.

Now, in order to prove
invariant $\gamma_1$ \cite[page 334]{pclentcs} for basic sequence $\BS_2$, we must
prove that $m$ is either generated, or received earlier as the main
message, by the same thread that sends out the signature:
\begin{multline*}
\gamma_1 \equiv 
\bigl( \psend(Y,t) \land
\pcontains(t,\tsig_{\hy}\crypt{y,m,\hx}) \bigr) \supset \\
\Bigl( \pgen(Y,m) \lor 
\bigl( \preceive(Y,(\hx,\hy,m)) < \ifacmver{ {} \\ }{}
    \psend(Y,(\hy,\hx,y,\tsig_{\hy}\crypt{y,m,\hx}))
    \bigr) \Bigr)
\end{multline*}\label{pg:gamma1}%
In order to prove this invariant with respect to basic sequence $\BS_2$,
one needs to
prove that $m$ is generated in the thread $Y$. However, there is no
mechanism to reason about any preceding actions, and thus it has to be
dealt with in the same way for both protocols $Q_{\CR}$ and $Q'$. One must
therefore also consider the case $m$ was {\em first received}, and not
generated, in a previous basic sequence as part of a bigger term, as
happens in $Q'$. In fact, the invariant is false for protocol $Q'$.

Thus, in order to prove the invariant for $Q_{\CR}$, 
we must be able to 
distinguish between the different semantics of $\BS_2$ within the context
of $Q_{\CR}$ and $Q'$.  Such
reasoning is not supported by basic PCL. 

Without the ability to reason about preceding actions, the protocol
descriptions are effectively cut up into independent basic sequences, which can
then only be dealt with as separate protocols. This is evident from the
formulation of the honesty rule \cite[page 329]{pclentcs}, in which the reference to the
protocol $Q$ is only used for the definition of the basic sequences. 
Thus one is required to create
much stronger proofs than would be strictly necessary. Put differently,
in basic PCL
one has to reason with an over approximation of protocol execution,
in which basic sequences occur in no fixed order in a thread.

This phenomenon can also be observed in the following example. Let $P$ be a
protocol, and let $P_1$ be the protocol
defined as $P$ extended with a role consisting of the
basic sequences $ \BS ; \BS' ; \BS''$. Similarly, let $P_2$ be defined as $P$ extended with a role $ \BS'' ; \BS' ; \BS$,
i.e., a role with the same basic sequences as the role in $P_1$, but
composed in a different order.
Then, any invariant $\gamma$ that can
be proven for protocol $P_1$ using the honesty rule, must also hold for $P_2$.
In fact, the proof is identical. Conversely, any invariant that holds for $P_1$
but not for $P_2$, cannot be proven using the honesty rule. This puts a strong
restriction on the type of invariants provable in basic PCL, because the structure
(and hence the properties) of both protocols can be very different.

\ifacmver{
\input{counterexample}
}{}

Ultimately, this problem seems to be a side effect of the weak link
between the protocol description $Q$ and the actions of honest agents in
a run $R \in \runsof(Q)$ in the protocol logic.  In basic PCL, the
only way to make the link between a protocol
description and the actions of honest agents, is by means of defining an
invariant and then proving it with the honesty rule. This can be seen
from inspecting the protocol logic rules: the only occurrences of a
protocol name (e.g.~$Q$) are in the honesty rule and the composition rules. As
the honesty rule only reasons about isolated basic sequences of the protocol,
the relations among the basic sequences of the protocol are inevitably lost.

\subsubsection{Resolving the problem}

Based on the semantics of PCL, it should be possible to extend the logic with a ``precedence'' inference rule, that
would allow one  to infer that given the $n$th action of a role $\rho$
occurs in a thread $X$, also the $(n-1)$th action of the same role must
have previously occurred in the same thread. 
Such an inference rule would allow for establishing preceding actions that must have occurred
from the existence of other actions. This is particularly useful for proving
invariants for basic sequences.

In general one could extend basic PCL with additional mechanisms to enable reasoning
about the relation between protocol descriptions and the actions of
honest agents in a run. However, the current weak link between the two
has one major advantage: it eases compositional proofs. For example,
because no constraints are put on any other (i.e.~non-protocol) actions of
honest agents, the sequential composition of protocols allows for re-use of proofs of
invariants for basic sequences. Thus, one must be careful when
introducing such links and introduce
only links between the protocol and the actions of honest agent, such
that the links are invariant under e.g.~sequential composition. A
``precedence'' inference rule of the type sketched above should meet
this condition.

\subsection{No formal type system}
\label{sec:notypes}

It is mentioned that in PCL, ``We assume enough typing to distinguish the keys
$K$ from the principals \myhat{A}, the nonces $n$ and so on.'' \cite[page 316]{pclentcs}.
However, there are no constructs in PCL that allow for formal reasoning about
the type of variables.

As a consequence, some claimed proofs of invariants are not
correct within the logic, as they are only true under type assumptions,
that cannot be expressed in PCL.

\subsubsection{Implications}

Many protocols have properties in a typed model, that do not hold for an
untyped model. In particular, some protocols are correct in a typed model,
but are vulnerable to so-called type flaw attacks in an untyped model.
Such attacks exploit for example that an agent could mistake an
encrypted session key for an encrypted nonce, and sends the key
unencrypted in a next step, revealing it to the intruder.
It is therefore often easier to prove properties for the typed model, but
this requires that the logic supports drawing conclusions based on 
types.  Currently, this is not possible in PCL.

As an example, we show that invariant 
$\gamma_1$ from \cite[page 334]{pclentcs}
(reproduced in this paper \myref{pg:gamma1}) is false when
variables are untyped.
The invariant $\gamma_1$ states
that if a thread (of an honest agent) sends a message that contains a signature
term $S$, with a subterm $m$, then either %
\newcommand{\gammaonea}{$m$ was generated in the thread $Y$}
\newcommand{\gammaoneb}{the thread $Y$ executes a \areceive of $m$, and later a \asend of the message tuple $(y,S)$}
\ifsaves{(1) \gammaonea, or (2) \gammaoneb. }{
\begin{enumerate}

  \item \gammaonea, or 

  \item \gammaoneb.

\end{enumerate}%
}%
Now consider the basic sequence $\BS_3$ from \cite[page
335]{pclentcs}:
\begin{align*}
\BS_3 &\equiv [ \; \areceive \hx,\hy,x ;\;
\anew n ;\;
r := \asign(n,x,\hx), \hy ;\;
\ifacmver{\\ & \quad \; \;}{}
\; \, \asend \hy,\hx,n,r ;\; ]_Y
\end{align*}\label{pg:bs3}%
This basic sequence corresponds to the responder receiving an unknown value,
supposedly a nonce, and sending back a signed version that includes a freshly
generated nonce.

In order to show the invariant is false, consider a thread $Y'$ that is executed
by agent $\hy$.
If we assume that $x = \tsig_{\hy}\crypt{y,m,\hx}$, where $m$ is generated by
thread $Y'$, the invariant is violated: 
we have that
$
  x = \tsig_{\hy}\crypt{y,m,\hx}
  \;
  \supset
  \;
  \pcontains(x,m)
$
and by substitution in $\BS_3$ and the $\pcontains$ predicate, we find that
$
  \pcontains(r, m)
$.
As a result, the message sent at the end of $\BS_3$ of a thread $Y$ ($Y'
\not= Y$) will contain $m$.
However, $m$ is neither generated by this thread nor is
it the exact term that was received. Thus the invariant $\gamma_1$ does
not hold for basic sequence $\BS_3$. Hence the example authentication proof of the $\CR$ protocol
in \cite{pclentcs} is incorrect.

\subsubsection{Resolving the problem}

The model can be extended with typing information for the variables,
and an axiom could be introduced that captures the fact that variables are only
instantiated by their typing information.  This would not introduce much new
machinery, but requires additional reasoning steps in most of the
proofs.

\section{Problems with PCL extensions}
\label{sec:pclextensions}

In this Section we discuss two additional mechanisms for PCL, in particular the extension
for Diffie-Hellman exponentials as found in \basicpcl and the extension for
	hash functions from \cite{thesis:he,pcltls}.

\subsection{Diffie-Hellman exponentials}

\label{sec:dh}

In basic PCL \basicpcl, axioms are provided for reasoning about
Diffie-Hellman (DH) exponentials. To that end, the logic is extended with
four additional axioms and some changes are made to the language and execution
model.

\subsubsection{Capturing Diffie-Hellman behaviour}

Below we recall the three elements of the DH
extension (execution model, logic, proof system) and discuss their
implications.

\ifsaves{}{
\mypar{DH extension of the language and execution model}
}

The programming language and execution model are extended \cite[page
354]{pclentcs} with
constructs $g(a)$ and $h(a,b)$, representing respectively $g^a \bmod p$
and $(g^{a} \bmod p)^{b} \bmod p$. In the PCL papers, the mod $p$ and
brackets are usually omitted, resulting in the abbreviations
$g^a$ and $g^{ab}$.

This extension does not reflect the actual
algebraic properties of the exponential. For Diffie-Hellman, the crucial
property is that $g^{ab} = g^{ba}$. This equation must be
included in the execution model: if it is not, the following sequence of
actions (denoted by their action predicates), which is perfectly valid
DH-type behaviour, does not correspond to a valid execution in PCL:
\begin{equation}
  \psend(X,h(a,b)) < \preceive(Y,h(b,a))
  \label{eq:dhactions}
\end{equation}
Just extending the $\phas$ predicate in the logic is not sufficient, as
the equivalence still has no counterpart in the execution model of PCL.

\ifsaves{}{
\mypar{DH extension of the protocol logic}
}

In basic PCL, the predicate $\pfresh(X,x)$ holds for any $x$ generated
by thread $X$ (captured by $\pgen(X,x)$), as long as $x$ is not sent as
part of another term.
The protocol logic is extended with an additional rule for the $\pfresh$ predicate,
stating that $\pfresh(X,g(x))$ is true if and only if $\pfresh(X,x)$ is true,
and $\phas$ is extended in \cite[page 354]{pclentcs} by
\[
  \phas(X,a) \land \phas(X,g(b)) \supset \phas(X,h(a,b)) \land \phas(X,h(b,a))
\]
This rule is intended to capture the Diffie-Hellman equivalence
relation. It is not sufficient, even with the addition of further
axioms, as we show below.

\ifsaves{}{
\mypar{DH extension of the proof system}
}

The proof system is extended in \cite[page
354, Table A.1]{pclentcs} by a definition and four new axioms. We
reproduce them here:
\begin{multline*}
\mbox{Define {\bf Computes (DH)}} \qquad \\
\pcomputes(X,g^{ab}) \equiv 
\Bigl(
\bigl(\phas(X,a) \land \phas(X,g^b)\bigr)
\lor
\ifacmver{ {} \\}{}
\bigl(\phas(X,b) \land \phas(X,g^a)\bigr)
\Bigr)
\end{multline*}
This definition captures the intuition that a thread can compute the
value of $g^{ab}$ if and only if it has the required components. Note that
for a thread that just has $g^{ab}$ but not $a$ or $b$, this
predicate is false, and therefore we have that $\phas(X,g^{ab})$ does {\em
not} imply $\pcomputes(X,g^{ab})$. 

Note that the definition of $\pcomputes$ suggests that 
$g^{ab} \not= g^{ba}$, because of the explicit listing of both sides of
the disjunction. If the equality would hold, one could just define 
$\pcomputes(X,g^{ab}) \equiv 
\phas(X,a) \land \phas(X,g^b)
$ to achieve the same result.

Using the definition of $\pcomputes$, the first axiom is given as:
\begin{multline*}
\shoveright{
\mbox{\bf DH1} \qquad 
\pcomputes(X,g^{ab})
\supset 
\phas(X,g^{ab})
}
\end{multline*}
and corresponds to the extension of the $\phas$ predicate, which
can be seen by unfolding the definition of $\pcomputes$. The second
axiom states
\newcommand{\dhtwopre}{\phas(X,g^{ab})}
\newcommand{\dhtwopost}{%
\Bigl(%
  \pcomputes(X,g^{ab}) \lor
  \ifacmver{ {} \\}{}
  \exists m.\bigl(
    \preceive(X,m) \land
    \pcontains(m,g^{ab})
  \bigr)%
\Bigr)%
}
\begin{multline*}
\ifsmall{%
\shoveright{\mbox{\bf DH2} \qquad \dhtwopre \supset { } 
\qquad \qquad \qquad \qquad \qquad \qquad
\qquad \qquad \qquad \qquad \qquad { }
}\\
\shoveleft{\qquad \hfill \dhtwopost}%
}{%
\ifacmver{
\mbox{\bf DH2} \qquad \dhtwopre \supset \dhtwopost 
}{%
\shoveright{ \mbox{\bf DH2} \qquad \dhtwopre \supset 
\dhtwopost }
}
}
\end{multline*}
This axiom captures the notion that one can only possess such a term
$g^{ab}$ by computing it from its parts, or receiving it. Effectively
this restricts any agent (including the intruder) from knowing a term of
the form $g^{ab}$ at the start of each run, but it also excludes that
$g^{ab}$ is used as a parameter for a protocol.

For completeness, we also give the remaining DH axioms
\ifacmver{
\begin{multline*}
\mbox{\bf DH3} \quad
\bigl( \preceive(X,m) \land \pcontains(m,g^{ab}) \bigr)
\supset \\
\exists Y, m'. \bigl(
  \pcomputes(Y,g^{ab}) \land
    \psend(Y,m') \land {} \\
    \pcontains(m',g^{ab})
  \bigr) 
\end{multline*}
\begin{multline*}
\shoveright{
\mbox{\bf DH4} \qquad 
\pfresh(X,a)
\supset 
\pfresh(X,g^a) 
}
\end{multline*}
}{%
\begin{align*}
\mbox{\bf DH3} \qquad &
\bigl( \preceive(X,m) \land \pcontains(m,g^{ab}) \bigr)
\supset \\
&
\qquad
\exists Y, m'. \bigl(
  \pcomputes(Y,g^{ab}) \land
    \psend(Y,m') \land
    \ifacmver{ {} \\ & \qquad \qquad
    }{}
    \pcontains(m',g^{ab})
  \bigr) 
  \qquad \qquad \qquad \qquad
  \\
\mbox{\bf DH4} \qquad 
& \pfresh(X,a)
\supset 
\pfresh(X,g^a) 
\end{align*}
}
which capture the assumption that $g^{ab}$ must have been either
computed or received (but was not given as a parameter). The fourth
axiom echoes the extension of the $\pfresh$ predicate.

Summarizing, even with the extension of the $\phas$ predicate and the
additional axioms, the behaviour of the equivalence for
Diffie-Hellman is not captured in PCL, for two reasons. First, with
respect to the execution model, the sequence of actions represented by
Formula~(\ref{eq:dhactions}) cannot be enabled in the execution model by
just changing the protocol logic. Second, we observe that given 
$\phas(X,h(a,b))$, the protocol logic
does not allow us to conclude that $\phas(X,h(b,a))$ as one would expect
(especially in the case where $\neg \pcomputes(X,h(a,b))$).
Hence, the essential equivalence for any Diffie-Hellman type protocol is
currently not captured in the execution model, nor in the protocol
logic.

\subsubsection{Implications}

The Diffie-Hellman extension does not correctly capture the algebraic
properties of Diffie-Hellman exponentials. As a result, certain possible
behaviours of DH-like protocols are not considered within the logic and
its execution model. The extension is therefore not a faithful
representation of DH-like protocols.

\subsubsection{Resolving the problem}

The essential feature to be captured is the equality $g^{ab} = g^{ba}$.
If this is introduced at the term level, e.g.~by having the equality
$h(a,b) = h(b,a)$ in the term system, this solves at the same time the
problem in the execution model, protocol logic and proof system. 
Some further modifications to the axioms are required, and the current
proofs have to be modified to take the term equality into account.

\subsection{Keyed hashes}

\label{sec:hashaxioms}

In \cite{thesis:he} and \cite{pcltls} basic PCL is extended and applied to a case study of protocols
that do not rely solely on signatures. As observed previously, such an
application
requires additional axioms. In particular, to prove authentication, we
need to introduce an axiom that allows us to conclude the existence of a
thread.

Appendix A of \cite{pcltls} gives such
additional axioms and definitions for keyed hash functions: one definition and four
hashing axioms. We reproduce them
here for convenience. 
\newcommand{\defhascomputes}{\pcomputes(X,\thash_K(a)) 
\equiv \phas(X,K) \land \phas(X,a) 
}%
\ifsmall{%
\begin{multline*}
\mbox{Define {\bf Computes (HASH)}} :\\
\defhascomputes
\end{multline*}
}{%
\ifacmver{%
\begin{multline*}
\mbox{Define {\bf Computes (HASH)}} :\\
\defhascomputes
\end{multline*}
}{%
\begin{equation*}
\mbox{Define {\bf Computes (HASH)}} \qquad \defhascomputes
\end{equation*}
}%
}%
As we will see in the following, the intention of this predicate seems
to be to express which thread computed the hashed value from its
components. However, this intuition is not correctly captured by the definition: a
typical use pattern of a keyed hash would be to provide an
integrity check for a message $m$, as in
\begin{equation}
  \hx \rightarrow \hy 
  \; : \;
  m,\thash_K(m) \mbox{ ,}
\end{equation}
where $K$ is a key shared between $\hx$ and $\hy$. In the typical use
pattern, the $\thash_K(m)$ is received by an agent which at that point can
construct the message $m$ himself. In this use case, the message hash is
received with (or after) the message, and is then used to verify the integrity
of the message.

Assume we
have that $m$ (or some subterm of it) is freshly generated by $X$, and
we have that $\pcomputes(X,\thash_K(m))$ holds. However, we can also prove
that once a thread $Y$ of the recipient $\hy$ receives the message, $Y$ not only has $K$, but also $m$.
Thus the predicate $\pcomputes(Y,\thash_K(m))$ holds as well.  Put
differently, we have that for a typical use case, $\pcomputes(\ldots)$ holds for
both the thread who creates as well as the thread who receives. In
particular, this is also true for the protocols under consideration in
\cite{pcltls}. The protocols in the paper assume typing information (no hash
mistaken for a nonce) and receive the hashed values at a point where
they know the hash components. 

Let $K$ be a
key shared between $\hx$ and $\hy$. Then, we 
can prove\footnote{Note that this cannot be proven using only 
basic PCL, but it can be proven using PCL combined with the
meta-reasoning required to capture the assumption that variables are
typed, as pointed out in Section~\ref{sec:notypes}.} the following
property (invariant) for the protocols in \cite{pcltls}:
\ifacmver{%
\begin{multline}
  \label{eq:computecollapse}
  \Honest(\hx) 
  \land
  \phas(X,\thash_K(m)) 
  \supset {} \\
  \pcomputes(X,\thash_K(m))
\end{multline}
}{%
\begin{equation}
  \label{eq:computecollapse}
  \Honest(\hx) 
  \land
  \phas(X,\thash_K(m)) 
  \supset
  \pcomputes(X,\thash_K(m))
\end{equation}
}
Intuitively one can see that an honest agent either creates the hash by
having the components, or receives it. In this last case, where an
honest agent receives the hash, the message integrity is verified using
the hash, which is possible only because the recipient also has both the key
$K$ and the message $m$.
We will use this result in our discussion of the {\bf HASH4} axiom later
on.

The first axiom in \cite[Appendix A]{pcltls} that uses the definition of $\pcomputes()$
is the following:
\begin{multline*}
\shoveright{
\mbox{\bf HASH1} \ifacmver{\\ \qquad}{\qquad }
\pcomputes(X,\thash_K(x))
\supset 
\phas(X,x) \land \phas(X,K)
}
\end{multline*}
If we unfold the definition of $\pcomputes$ in this axiom, it
reduces to $\phi \supset \phi$.
The second hash axiom states the following:
\begin{multline*}
\shoveright{
\mbox{\bf HASH2} \ifacmver{\\ \qquad}{\qquad }
\pcomputes(X,\thash_K(x))
\supset
\phas(X,\thash_K(x))
}
\end{multline*}
This axiom informally states that whoever computes the hash value also has
it. If we again unfold the definition of $\pcomputes$, we can see that
this works as an extension of the closure properties of the $\phas$
predicate (as defined in \cite{pclentcs}). Effectively, it introduces a
new term structure $f_x(y)$ to the PCL syntax, and expresses the one-way
property of the hash function.

The third hash function axiom is
\newcommand{\hashthreepre}{ \preceive(X, \thash_K(x)) }
\newcommand{\hashthreepost}{ \exists Y. \pcomputes(Y,\thash_K(x)) 
\land \psend(Y,\thash_K(x))
\mbox{ .}
}
\ifsmall{%
\begin{multline*}
\mbox{\bf HASH3} \quad 
\hashthreepre \supset \\
\hashthreepost
\end{multline*}
}{%
\ifacmver{%
\begin{multline*}
\mbox{\bf HASH3} \quad 
\hashthreepre \supset \\
\hashthreepost
\end{multline*}
}{%
\begin{equation*}
\mbox{\bf HASH3} \quad
\hashthreepre \supset \hashthreepost
\end{equation*}
}
}
This axiom is not sound. 
Consider the following protocol in
Alice and Bob style notation,
where $m$ is a freshly generated nonce of the initiator, and $K'$
is a symmetrical key shared by the initiator and responder:
\begin{equation}
\begin{split}
  Init \rightarrow Resp & \; : \; \tenc_{K'}\crypt{\thash_K(m)} \\
  Resp \rightarrow Init & \; : \; \thash_K(m) 
\end{split}
\label{prot:hash3}
\end{equation}
In a normal run of this protocol, the initiator sends the hash as part
of a bigger term, but does not send $m$. Thus the responder cannot
compute the hash itself, but simply decrypts the message, and sends the
hash back.
Thus, the preconditions are fulfilled by an initiator thread of this
protocol, but the postcondition does not hold: only the initiator thread
can compute it, but it does not send it out in the required form.
(Observe that contrary to the protocols in \cite{pcltls}, this protocol
in~(\ref{prot:hash3}) does not satisfy the typical usage pattern, and therefore
Formula~(\ref{eq:computecollapse}) does not hold here.)

The fourth axiom is
\begin{multline*}
\mbox{\bf HASH4} \ifacmver{\;}{\quad}
\phas(X,\thash_K(x))
\ifacmver{\!}{}
\supset 
\ifacmver{\!}{}
\pcomputes(X,\thash_K(x)) 
\ifacmver{ {}\\}{}
\lor \ifacmver{}{ {} \\}
\exists Y,m. \; \pcomputes(Y,\thash_K(x))
\land
\psend(Y,m)
\land
\ifacmver{ {}\\}{}
\pcontains(m,\thash_K(x))
\mbox{ .}
\end{multline*}
This axiom seems to express: if someone has the hash value, she computed
the hash herself, or somebody computed it and sent it (possibly as a
subterm). 

With a suitable restriction on the intruder knowledge
(the intruder should not know some $\thash_K(x)$ initially without knowing the components)
this axiom can be proven sound.
However, there is a problem with the
applications of the axiom in the proofs of \cite{pclentcs}. Each time this axiom is applied in proofs in
the paper, one assumes honesty of $\hx$ and $\hy$, and that
$K$ is the key shared between them. Thus we can rewrite the application
of axiom {\bf HASH4} using Formula~(\ref{eq:computecollapse}), as in the
following. First we unfold the definition of the axiom
\begin{multline}
\Honest(\hx) 
\land 
{\bf HASH4}
= \\
\Honest(\hx) 
\land
\bigl(
\phas(X,\thash_K(x))
\supset \ifacmver{ {}\\}{}
\pcomputes(X,\thash_K(x)) 
\lor \Phi
\bigr) \mbox{ ,}
\end{multline}
where $\Phi$ is used to denote the right-hand disjunct of the axiom {\bf HASH4},
starting from the existential quantifier.  Now we use
Formula~(\ref{eq:computecollapse}) to get
\begin{equation}
\Honest(\hx) 
\land 
{\bf HASH4}
=
\Honest(\hx) 
\mbox{ .}
\end{equation}
In other words, if $\hx$ is assumed to be honest when applying
axiom {\bf HASH4}, the left-hand disjunct of the conclusion is always
true (rendering the right-hand disjunct inconsequential). Because the
right-hand disjunct of the conclusion contains the required thread
existence, but is rendered useless, the axiom cannot be used here for
proving authentication properties, based on our observations in
Section~\ref{sec:authexistence}.

\subsubsection{Implications}

The authentication proofs of the four-way handshake and group key handshake
protocols in \cite{pcltls,thesis:he} are not correct in their current form. The
reason for this is that these protocols do not contain signatures, and
based on the observations in Section~\ref{sec:authexistence}, any
authentication proofs for such protocols must rely on newly introduced
axioms.
In this case, the only candidates 
are the axioms {\bf HASH3} and {\bf HASH4}. As shown above, {\bf HASH3} is not
sound, and for the protocols in \cite{pcltls,thesis:he}, {\bf HASH4} cannot be
used to prove the existence of a thread. Hence the authentication proofs of the handshakes are incorrect, and therefore also the compositional proof
of authentication of IEEE 802.11i is incorrect.

\subsubsection{Resolving the problem}

The properties of a keyed hash function like the ones occurring here are
similar to the properties of symmetric-key cryptography. Thus,
once sufficient reasoning infrastructure is in place for symmetric-key
cryptography, one can devise a straightforward definition of a
non-keyed hash function. Combining these two elements should lead to a
natural definition and extension of the logic for keyed hashes.

Alternatively, it might be possible to reinstate rules that were used in older
works, as e.g.~found in \cite[page 15]{csfw04}, assuming that their
soundness can be proven in the current semantics.

\section{Conclusions}
\label{sec:conclusions}

In this paper, we have first shown that basic PCL as defined
in \basicpcl cannot be used to prove authentication
properties of protocols without signatures. We consider this to be a
strong limitation on the scope of basic PCL.
Next, we have pointed out two reasoning tools that are missing in basic PCL%
:
reasoning about preceding actions of a role within a thread, and the lack of a formal
type system. 
With respect to PCL extensions, we have shown that the PCL
Diffie-Hellman extension from \basicpcl does not capture the algebraic behaviour of
Diffie-Hellman protocols correctly in the execution model and protocol
logic.  Finally, we have shown that the
extension for protocols with keyed hash functions
in \cite{pcltls,thesis:he} is
not sound.  

It follows that some currently claimed PCL proofs
cannot be proven within the basic logic as defined in \basicpcl, or make use of unsound axioms.
These proofs include the authentication proofs of the $\CR$ protocol from
\cite{pclentcs,pcljournal}
and
the SSL/TLS and IEEE 802.11i protocols from \cite{thesis:he,pcltls}%
.

In the
papers on PCL the proofs of the invariants are often not given. This lack
of proof details for the invariants is surprising, as the invariants themselves
are the difficult part. Furthermore, many of
the proofs seem impossible to be completed in PCL without resorting to
meta-reasoning.

Some of the problems identified here can be solved easily, but for some
problems more work is required. For example, straightforward solutions include adding a
formal type system and adding a mechanism to reason about earlier
actions. Other problems,
such as establishing thread existence of honest agents for protocols based on public-key
cryptography, do not seem to be solvable by straightforward fixes, and suggest
that more extensive modifications to PCL are required.

It remains to be
seen whether formal proofs in such a modified version of PCL can be concise.


\bibliographystyle{alpha}
\bibliography{biblio}

\end{document}
